\documentstyle[12pt,cite]{article}

  \def\half{{1 \over 2}}
  \def\i3{{1 \over 3}}
  \def\sqr2{\sqrt{2}}
  \def\isqr2{{1 \over \sqrt{2}}}

  \def\tO{\tilde{\Omega}}

  \def\to{\tilde{\omega}}
  \def\tv{\tilde{\mbox {\boldmath $v$}}}

\begin{document}

\vspace*{0.3cm}
\begin{Large}
\begin{center}
Universal Seesaw Mechanism with Universal Strength\\
for Yukawa Couplings
\end{center}
\end{Large}

\vskip 0.5cm
\begin{large}
\begin{center}
Tadatomi {\sc Shinohara}$^{a,}$\footnote{E-mail: sinohara@cc.kyoto-su.ac.jp}, 
Hajime {\sc Tanaka}$^{a,}$\footnote{E-mail: gen@cc.kyoto-su.ac.jp} \\
and Ikuo S. {\sc Sogami}$^{a,b,}$\footnote{E-mail: sogami@cc.kyoto-su.ac.jp} \\
\vskip 0.3cm
{\it 
${}^a$\ Department of Physics, Kyoto Sangyo University, \\
        Kyoto 603-8555, Japan\\
${}^b$\ Theoretical Physics Division, CERN, \\
        CH-1211 Geneva 23, Switzerland}
\end{center}
\end{large}

\vskip 1.0cm
\begin{abstract}
Hypotheses of the universal seesaw mechanism and the {\it universal strength
for Yukawa couplings} are applied to explain one possible
origin of quasi-democratic mass matrices of a special type in a left-right
symmetric model with the gauge
group $SU(3)_c\times SU(2)_L\times SU(2)_R\times U(1)$.
Two kinds of Higgs doublets are postulated to mediate scalar interactions
between the $i$-th generation of light fermion doublets and the $j$-th
generation of heavy fermion singlets with relative Yukawa coupling constants
of the exponential form $e^{i\phi_{ij}}$, where $\phi_{ij}$ are real phase
constants.  The lowest seesaw approximation results effectively in
self-adjoint mass matrices which are quasi-democratic and have the same
diagonal elements.  A set of values for the parameters $\phi_{ij}$ is found
which reproduces the present experimental data for the absolute values of
the CKM matrix elements, the Jarlskog parameter and the Wolfenstein
parameters.
\end{abstract}

\newpage

\section{Introduction}

Among various forms of mass matrices\cite{rf:1,rf:2,rf:3,rf:4,rf:5,rf:6},
quasi-democratic mass matrices explain hierarchical structures of
quark mass spectra and the CKM weak-mixing matrix in a simple and systematic
way\cite{rf:7,rf:8,rf:9,rf:10,rf:11,rf:12,rf:13,rf:14,rf:15,rf:16,rf:17,
rf:18,rf:19,rf:20,rf:21}. In a previous article\cite{rf:21} we investigated
the eigenvalue problem of quasi-democratic mass matrices of special type,
 \begin{equation}
   {\cal M}_q = M_q \tO_q ,  \qquad (q = u, d)
 \label{massmatrix1}
 \end{equation}
where $M_q$ is a mass scale for the q-sector and $\tO_q$ is the Hermitian
matrix
 \begin{equation}
   \tO_q = {1 \over 3}
   \left( \matrix{1 & a_3^q \, e^{ i\delta_{12}^q} &
                      a_2^q \, e^{-i\delta_{31}^q}   \cr
                  a_3^q \, e^{-i\delta_{12}^q} & 1 &
                  a_1^q \, e^{ i\delta_{23}^q}       \cr
                  a_2^q \, e^{ i\delta_{31}^q} &
                  a_1^q \, e^{-i\delta_{23}^q} & 1   \cr}
   \right)
 \label{massmatrix2}
 \end{equation}
with phases satisfying the restriction
 \begin{equation}
   \delta_{12}^q = 0, \quad \delta_{23}^q = -\delta_{31}^q = \phi_q.
 \end{equation}
Solving the mass eigenvalue problem in the first order perturbation
approximation with respect to small deviations around the democratic
limit $a_j^q =1$ and $\phi_q = 0$, we found sum rules for the absolute values
of the CKM matrix\cite{rf:22} elements and approximate expressions for the
Jarlskog parameter\cite{rf:23} and the Wolfenstein parameters\cite{rf:24}, all
of which are consistent with the present experimental data. The purpose of
this article is to derive effectively the above specific mass matrices from
the hypothesis of the {\it universal strength for Yukawa couplings} (USY)
through the universal seesaw mechanism.

Branco, Silva and Rebelo\cite{rf:15} first formulated the hypothesis of
the USY which results in pure phase mass matrices of the form
 \begin{equation}
   {\cal M}_q = c_q \Bigl[ e^{i\phi_{ij}^q} \Bigr] ,  \qquad
   (q = u, d)
 \end{equation}
and there followed many similar investigations\cite{rf:16,rf:17,rf:18,rf:19}.
Fishbane and Hung\cite{rf:19} gave an example of Higgs field interactions
that can produce the pure phase mass matrices.
Since these pure phase mass matrices are not necessarily Hermitian,
it is usual to introduce the dimensionless Hermitian matrices
$H_q = {1 \over 3{c_q^2}} {\cal M}_q {\cal M}_q^\dagger$ and solve the
eigenvalue problem for $H_q$. Note that our quasi-democratic mass matrices
$\tO_q$ have the same form as the Hermitian matrices $H_q$.

The universal seesaw mechanism (USM)\cite{rf:25,rf:26,rf:27,rf:28,rf:29,rf:30}
was invented to explain the smallness of the charged fermion masses relative
to the electroweak scale by postulating the existence of exotic fermions
belonging to electroweak singlets.  In Ref.~\cite{rf:28} we considered
the model based on the left-right-symmetric gauge group\cite{rf:31}
 \begin{displaymath}
   G \equiv SU(3)_c \times SU(2)_L \times SU(2)_R \times U(1)_y ,
 \end{displaymath}
where a group $U(1)_y$ is generated by a new charge $y$. In the model,
the chiral charges are assigned to the fermion and Higgs fields
to distinguish generations and to restrict the pattern of the Yukawa
interactions leading to mass matrices of the extended Fritzsch type\cite{rf:5}
in the lowest seesaw approximation.

In this article the universal seesaw mechanism is implemented with the above
mentioned left-right symmetric gauge group $G$\cite{rf:31}.  We choose the
simplest Higgs structure with a pair of Higgs doublets $\chi_L$ and $\chi_R$
and postulate that these two doublets interact with the fermions with
the universal strength $Y$ of the Yukawa couplings. Namely, the Higgs fields
are assumed to mediate scalar interactions between the $i$-th generation of
the light fermion doublets and the $j$-th generation of the heavy fermion
singlets with Yukawa coupling constants of form $Ye^{i\phi_{ij}}$, where
the $\phi_{ij}$ are real phase constants. Note here that the universality
of the strength for Yukawa couplings is formulated more stringently than that
of Branco~et~al.\cite{rf:15}, since $Y$ is common to the up- and
down-quark sectors.  The mass $m_Q\ (Q=U,D)$ is assigned to the up- and
down-heavy fermion singlets.
Through the breakdown of the symmetry $G$ the Higgs doublets $\chi_L$ and
$\chi_R$ acquire, respectively, the vacuum expectation values $w_L$ and $w_R$.
Under the first seesaw approximation, effective Hermitian mass matrices of
the quasi-democratic form in Eqs.~(\ref{massmatrix1}) and (\ref{massmatrix2})
are obtained.  To take the first seesaw approximation corresponds just
to the product operation ${\cal M}_q {\cal M}_q^\dagger$, as shown by
Branco~et~al.  The mass scale $M_q$ in Eq.~(\ref{massmatrix1}) is determined by
 \begin{equation}
   M_q = m_1^q + m_2^q + m_3^q \simeq 9Y^2 {w_L w_R \over m_Q}.
 \end{equation}
Thus the scale difference for the up- and down-quark sectors, $M_u$ and $M_d$,
is reduced to that of heavy quark masses, $m_U$ and $m_D$. The departure
of $M_u$ and $M_d$ from the electroweak scale $w_L$ is explained
by the factor $9Y^2 w_R / m_Q$.

Note that all results in the previous article\cite{rf:21} hold here,
since the effective mass matrices coincides with the quasi-democratic mass
matrix of the specific form in Eqs.~(\ref{massmatrix1}) and
(\ref{massmatrix2}).  Namely, the present model can inherit all the results
obtained there.

This article is organized as follows.
The model is formulated in {\S}2. We derive the seesaw mass matrix in {\S}3
and parametrize the effective mass matrices in {\S}4. Exact solutions of
the mass eigenvalue problems are given, and their relations with observable
quantities are explained in {\S}5. Results of numerical analyses are given
in {\S}6, and discussion is given in {\S}7.

\section{Model}

Fundamental quarks in the model are classified with respect to the underlying
gauge group $G$. The ordinary quarks belonging to the {\it i}-th
generation $(i = 1,2,3)$ have the transformation properties
 \begin{equation}
   q_{iL} = \left( \matrix{u^r & u^g & u^b \cr
                           d^r & d^g & d^b \cr} \right)_{iL}
       \sim (3, 2, 1; {1 \over 3}) ,
 \end{equation}
 \begin{equation}
   q_{iR} = \left( \matrix{u^r & u^g & u^b \cr
                           d^r & d^g & d^b \cr} \right)_{iR}
       \sim (3, 1, 2; {1 \over 3}) ,
 \end{equation}
where the fourth entry in the parentheses is the $y$ charge generating
$U(1)_y$ group.  To implement USM, electroweak singlets of
exotic quarks $U_i$ and $D_i$, which have chiral projections transforming as
 \begin{equation}
   U_{ih} = \left( \matrix{U^r & U^g & U^b \cr} \right)_{ih}
       \sim (3, 1, 1; {4 \over 3}) ,  \qquad (h = L, R)
 \end{equation}
 \begin{equation}
   D_{ih} = \left( \matrix{D^r & D^g & D^b \cr} \right)_{ih}
       \sim (3, 1, 1; -{2 \over 3})  \qquad (h = L, R)
 \end{equation}
are introduced as the seesaw partners for each generation.

We introduce $SU(2)_h\ (h = L, R)$ doublets of Higgs fields as
 \begin{equation}
   \chi_L \sim (1, 2, 1; -1), \qquad \chi_R \sim (1, 1, 2; -1),
 \end{equation}
\noindent
which develop the vacuum expectation values
 \begin{equation}
   \langle \chi_L \rangle = \left( \matrix{w_L \cr 0 \cr} \right), \qquad
   \langle \chi_R \rangle = \left( \matrix{w_R \cr 0 \cr} \right)
 \end{equation}
\noindent
to break the left-right symmetry and the Weinberg-Salam symmetry.
Here, $w_L$ and $w_R$ are assumed to be real.

Our hypothesis of the USY asserts that the Higgs fields mediate scalar
interactions between the $i$-th generation of the light fermion doublets
and the $j$-th generation of the heavy fermion singlets with Yukawa coupling
constants of the form $Ye^{i\phi_{ij}^f}\ (f = u, d)$.
The heavy fermion singlets $U_{ih}$
($D_{ih}$) are assumed to have a degenerate bare mass $m_U$ ($m_D$).
The most general form of the fermion Lagrangian density ${\cal L}_Y$
satisfying these requirements, and being invariant under
the fundamental group $G$ and the $L \leftrightarrow R$-symmetry, is given by
 \begin{eqnarray}
   {\cal L}_Y & = & Y \sum_{i,j} \{
                       e^{i\phi_{ij}^u} {\bar q}_{iL} \chi_L U_{jR}
                     + e^{i\phi_{ij}^d} {\bar q}_{iL} {\tilde \chi}_L D_{jR}
                     + e^{i\phi_{ij}^u} {\bar q}_{iR} \chi_R U_{jL}
                     + e^{i\phi_{ij}^d} {\bar q}_{iR} {\tilde \chi}_R U_{jL}
                                 \} \nonumber \\
              & & - \sum_i \{ m_U {\bar U}_{iL} U_{iR}
                            + m_D {\bar D}_{iL} D_{iR} \} + {\rm h.c.} ,
 \end{eqnarray}
\noindent
where $\tilde{\chi} = i \sigma_2 \chi^\ast$. It is worthwhile to emphasize
that $Y$ is the common strength of the Yukawa couplings for the
up- and down-quark sectors. The minus sign for the bare mass terms
of heavy quarks is introduced to adjust the sign of the effective quark
mass matrices $M^q_{\rm eff}$ in {\S}4.

\section{Seesaw mass matrices}

Spontaneous breakdowns of the underlying symmetry $G$ induce $6 \times 6$
seesaw mass matrices for the up- and down-quark sectors. The mass matrices
are expressed in seesaw block-matrix form as
 \begin{equation}
   \left( \matrix{{\bar f}_L & {\bar F}_L \cr} \right)
   \left( \matrix{  0   & M_L^f \cr
                  M_R^f &  M^F  \cr} \right)
   \left( \matrix{f_R \cr F_R \cr} \right) + {\rm h.c.} ,
 \end{equation}
where $f = (f_1, f_2, f_3)^{\rm T}$ are the column vectors of three ordinary
quarks $(f = u, d)$ in the generation space,
and $F = (F_1, F_2, F_3)^{\rm T}$ are the column vectors of exotic
quarks $(F = U, D)$.  The component mass matrices $M_L^f$ and
$M_R^f$ are written
 \begin{equation}
   M_L^f = w_L M^f, \quad M_R^f = w_R M^{f\dagger}
 \end{equation}
in terms of the pure phase matrix $M^f$ defined by
 \begin{equation}
   M^f \equiv Y
   \left( \matrix{e^{i\phi_{11}^f} & e^{i\phi_{12}^f} & e^{i\phi_{13}^f} \cr
                  e^{i\phi_{21}^f} & e^{i\phi_{22}^f} & e^{i\phi_{23}^f} \cr
                  e^{i\phi_{31}^f} & e^{i\phi_{32}^f} & e^{i\phi_{33}^f} \cr
                 } \right) .
 \label{phasematrix}
 \end{equation}
The submass matrix $M^F$ has the diagonal structure
 \begin{equation}
   M^F = - \left( \matrix{m_F &  0  &  0  \cr
                           0  & m_F &  0  \cr
                           0  &  0  & m_F \cr} \right)
       = - m_F E ,
 \end{equation}
\noindent
where $E$ is $3\times3$ unit matrix.

On the assumption that
 \begin{equation}
   m_Q^2 \gg Y^2 w_R^2 \gg Y^2 w_L^2 , \qquad (Q = U, D)
 \label{seesawcondition}
 \end{equation}
we can apply the first seesaw approximation by
ignoring ${\cal O}(Y^2 w_{L,R}^2/m_F^2)$ in comparison with 1.
Note that this seesaw condition is the special one induced by the hypothesis
of USY.  The seesaw mass matrices are block-diagonalized
approximately by the bi-unitary transformations as
 \begin{equation}
   U_L^{f\dagger} \left( \matrix{  0   & M_L^f \cr
                                 M_R^f &  M^F  \cr} \right) U_R^f \simeq
   \left( \matrix{M_{\rm eff}^f &      0        \cr
                        0       & M_{\rm eff}^F \cr} \right)
 \end{equation}
with
 \begin{equation}
   U_{L,R}^f = \left(
   \matrix{1 - {1 \over 2}\rho_{L,R}^{f\dagger} \rho_{L,R}^f &
            \rho_{L,R}^{f\dagger}                            \cr
           -\rho_{L,R}^f                                     &
           1 - {1 \over 2}\rho_{L,R}^f \rho_{L,R}^{f\dagger} \cr}
               \right) .
 \end{equation}
Here the component matrices $\rho_L^f$ and $\rho_R^f$ of the bi-unitary
transformations are given by
 \begin{equation}
   \rho_L^{f\dagger} = M_L^f (M^F)^{-1}, \quad
   \rho_R^f          = (M^F)^{-1} M_R^f .
 \end{equation}
Therefore,the effective mass matrices for the ordinary and exotic quarks,
$M_{\rm eff}^f$ and $M_{\rm eff}^F$, are obtained in the forms
 \begin{equation}
   M_{\rm eff}^f \equiv - M_L^f (M^F)^{-1} M_R^f , \quad
   M_{\rm eff}^F \equiv M^F .
 \end{equation}

\section{Parametrization of effective mass matrices}

Here let us use the suffices $q = u, d$ and $Q = U, D$, since our arguments
are restricted to quarks.

Owing to the specific structures of the component matrices $M_L^q$, $M_R^q$
and $M^Q$, the effective mass matrix $M_{\rm eff}^q$ for ordinary quarks
is expressed by the Hermitian matrix
 \begin{equation}
    M_{\rm eff}^q = - M_L^q (M^Q)^{-1} M_R^q =
    {w_L w_R \over m_Q} M^q M^{q\dagger}.
 \label{eq:401}
 \end{equation}
In order to cast the effective mass matrix into the specific form
in Eqs.~(\ref{massmatrix1}) and (\ref{massmatrix2}), we parametrize
the Hermitian matrix $M^q M^{q\dagger}$ as
 \begin{equation}
   M^q M^{q\dagger} \equiv
   3 Y^2 \left( \matrix{1 & a_3^q \, e^{ i\delta_{12}^q} &
                            a_2^q \, e^{-i\delta_{31}^q}   \cr
                        a_3^q \, e^{-i\delta_{12}^q} & 1 &
                        a_1^q \, e^{ i\delta_{23}^q}       \cr
                        a_2^q \, e^{ i\delta_{31}^q} &
                        a_1^q \, e^{-i\delta_{23}^q} & 1   \cr}
         \right) .
 \label{eq:402}
 \end{equation}
Here, $a_1^q$, $a_2^q$, $a_3^q$, $\delta_{12}^q$, $\delta_{31}^q$ 
and $\delta_{23}^q$ are real parameters which are expressed
in terms of the original phases $\phi_{ij}^q$ in Eq.~(\ref{phasematrix})
as follows:
 \begin{eqnarray}
   a_1^q & = & {1 \over 3} [3
           + 2\cos(\phi_{21}^q + \phi_{32}^q - \phi_{22}^q - \phi_{31}^q)
           + 2\cos(\phi_{21}^q + \phi_{33}^q - \phi_{31}^q - \phi_{23}^q)
                             \nonumber \\
         &   & {\hskip 1.5cm}
           + 2\cos(\phi_{32}^q + \phi_{23}^q - \phi_{22}^q - \phi_{33}^q)
                              ]^{1/2} ,
 \label{eq:403}
 \end{eqnarray}
 \begin{eqnarray}
   a_2^q & = & {1 \over 3} [3
           + 2\cos(\phi_{11}^q + \phi_{32}^q - \phi_{12}^q - \phi_{31}^q)
           + 2\cos(\phi_{11}^q + \phi_{33}^q - \phi_{31}^q - \phi_{13}^q)
                             \nonumber \\
         &   & {\hskip 1.5cm}
           + 2\cos(\phi_{32}^q + \phi_{13}^q - \phi_{12}^q - \phi_{33}^q)
                              ]^{1/2} ,
 \label{eq:404}
 \end{eqnarray}
 \begin{eqnarray}
   a_3^q & = & {1 \over 3} [3
           + 2\cos(\phi_{11}^q + \phi_{22}^q - \phi_{12}^q - \phi_{21}^q)
           + 2\cos(\phi_{11}^q + \phi_{23}^q - \phi_{21}^q - \phi_{13}^q)
                             \nonumber \\
         &   & {\hskip 1.5cm}
           + 2\cos(\phi_{22}^q + \phi_{13}^q - \phi_{12}^q - \phi_{23}^q)
                              ]^{1/2} ,
 \label{eq:405}
 \end{eqnarray}
 \begin{equation}
   \delta_{12}^q = \tan^{-1} \left[ \displaystyle{
         \sin(\phi_{11}^q - \phi_{21}^q) + \sin(\phi_{12}^q - \phi_{22}^q) +
         \sin(\phi_{13}^q - \phi_{23}^q) \over 
         \cos(\phi_{11}^q - \phi_{21}^q) + \cos(\phi_{12}^q - \phi_{22}^q) +
         \cos(\phi_{13}^q - \phi_{23}^q)
                                                  } \right] ,
 \label{eq:406}
 \end{equation}
 \begin{equation}
   -\delta_{31}^q = \tan^{-1} \left[ \displaystyle{
         \sin(\phi_{11}^q - \phi_{31}^q) + \sin(\phi_{12}^q - \phi_{32}^q) +
         \sin(\phi_{13}^q - \phi_{33}^q) \over 
         \cos(\phi_{11}^q - \phi_{31}^q) + \cos(\phi_{12}^q - \phi_{32}^q) +
         \cos(\phi_{13}^q - \phi_{33}^q)
                                                   } \right] ,
 \label{eq:407}
 \end{equation}
and
 \begin{equation}
   \delta_{23}^q = \tan^{-1} \left[ \displaystyle{
         \sin(\phi_{21}^q - \phi_{31}^q) + \sin(\phi_{22}^q - \phi_{32}^q) +
         \sin(\phi_{23}^q - \phi_{33}^q) \over 
         \cos(\phi_{21}^q - \phi_{31}^q) + \cos(\phi_{22}^q - \phi_{32}^q) +
         \cos(\phi_{23}^q - \phi_{33}^q)
                                                  } \right].
 \label{eq:408}
 \end{equation}

Note that, if the phases are redefined by making the replacement
$\phi_{ij}^q - \phi_{jj}^q \Rightarrow \phi_{ij}^q$, it is possible
to eliminate the terms $\phi_{ii}$ from all these expressions in
Eqs.~(\ref{eq:403}) $\sim$ (\ref{eq:408}). Therefore, without loss
of generality, we can set $\phi_{11}^q = \phi_{22}^q = \phi_{33}^q$ = 0.

\section{Mass eigenvalue problems and observable quantities}

In the manner described above the first seesaw approximation leads to
the effective mass matrices
 \begin{equation}
   M_{\rm eff}^q = M_q \tO_q ,
 \label{eq:501}
 \end{equation}
\noindent
where $\tO_q$ are the Hermitian matrices $\tO_q$ in Eq.~(\ref{massmatrix2}),
and the mass scales $M_q$ are fixed as
 \begin{equation}
   M_q = 9Y^2 {w_Lw_R \over m_Q}.
 \label{eq:502}
 \end{equation}
To solve the eigenvalue problems for effective mass matrices, it is convenient
to simplify $\tO_q$ further by the unitary (phase) transformation as
 \begin{equation}
     \tO_q' \equiv {\cal P}_q^\dagger {\tilde \Omega}_q {\cal P}_q
   = {1 \over 3}
     \left( \matrix{  1   & a_3^q &       a_2^q            \cr
                    a_3^q &   1   & a_1^q \, e^{i\Delta_q} \cr
                    a_2^q & a_1^q \, e^{-i\Delta_q} &   1  \cr}
                        \right) ,
 \label{eq:503}
 \end{equation}
with the unitary matrix given by
 \begin{equation}
     {\cal P}_q
   = \left( \matrix{1 &      0      &            0       \cr
                    0 & e^{-i\delta_{12}^q} &    0       \cr
                    0 &      0      & e^{i\delta_{31}^q} \cr}
                        \right) ,
 \label{eq:504}
 \end{equation}
\noindent
where $\Delta_q = \delta_{12}^q + \delta_{23}^q + \delta_{31}^q$.

The eigenvalue problems
 \begin{equation}
   \tO_q' \tv_j^q = \to_j^q \tv_j^q
 \label{eq:505}
 \end{equation}
are solved with the eigenvectors $\tv_j^q$ in the form
 \begin{equation}
   \tv_j^q = N_j^q
   \left( \matrix{e^{i\Delta_q} a_1^q a_3^q + (3\to_j^q - 1) a_2^q \cr
                  e^{i\Delta_q} (3\to_j^q - 1) a_1^q + a_2^q a_3^q \cr
                                (3\to_j^q - 1)^2 - (a_3^q)^2       \cr}
   \right) ,
 \label{eq:506}
 \end{equation}
\noindent
where $N_j^q$ are the normalization constants given by
 \begin{eqnarray}
   |N_j^q|^{-2} & = &
        3(3\to_j^q - 1)^4 -
        [4(a_3^q)^2+(a_2^q)^2+(a_1^q)^2] (3\to_j^q - 1)^2 \nonumber \\
               & &
      + (a_3^q)^2 [(a_3^q)^2 + (a_2^q)^2 + (a_1^q)^2] .
 \label{eq:507}
 \end{eqnarray}
Accordingly, the Hermitian matrix $\tO_q$ is diagonalized as
 \begin{equation}
   U_q^\dagger {\cal P}_q^\dagger \tO_q {\cal P}_q U_q =
   U_q^\dagger \tO_q' U_q =
   \left( \matrix{\to_1^q & 0 & 0 \cr
                  0 & \to_2^q & 0 \cr
                  0 & 0 & \to_3^q \cr} \right) ,
 \label{eq:508}
 \end{equation}
where
 \begin{equation}
   U_q = \left( \matrix{\tv_1^q & \tv_2^q & \tv_3^q \cr} \right) .
 \label{eq:509}
 \end{equation}
The eigenvalues $\to_j^q$ satisfy the following relations:
 \begin{equation}
 \left\{ \begin{array}{l}
   \to_1^q \to_2^q \to_3^q = \displaystyle{1 \over 27} \left[
   1 - (a_3^q)^2 - (a_2^q)^2 - (a_1^q)^2
     + 2 a_3^q a_2^q a_1^q \cos\Delta_q                \right] ,  \\
       \noalign{\vskip0.2cm}
   \to_1^q \to_2^q + \to_2^q \to_3^q + \to_3^q \to_1^q =
   \displaystyle{1 \over 9} \left[
   3 - (a_3^q)^2 - (a_2^q)^2 - (a_1^q)^2
                            \right] ,  \\
       \noalign{\vskip0.2cm}
   \to_1^q + \to_2^q + \to_3^q = 1 .
 \end{array} \right.
 \label{eq:510}
 \end{equation}

The masses $m_j^q$ of ordinary quarks and the eigenvalues $\to_j^q$ of the
Hermitian matrix $\tO_q$ are related by
 \begin{equation}
   m_j^q = M_q \to_j^q,  \qquad
   \to_j^q = {m_j^q \over m_1^q + m_2^q + m_3^q} .
 \label{eq:511}
 \end{equation}
Eliminating the eigenvalues $\to_j^q$ in Eqs.~(\ref{eq:502}) and
(\ref{eq:511}), we obtain
 \begin{equation}
   M_q =
   m_1^q + m_2^q + m_3^q = 9Y^2 {w_Lw_R \over m_Q} , \qquad
   {m_1^u + m_2^u + m_3^u \over m_1^d + m_2^d + m_3^d} =
   {m_D \over m_U} .
 \label{eq:512}
 \end{equation}
Therefore, the different mass scales in the up- and down-quark sectors
come from the difference of bare masses of the exotic quarks $U$ and $D$.
The gaps of the mass scales $M_u$ and $M_d$ from the electroweak scale $w_L$
are suppressed by the seesaw factor $9 Y^2 w_R / m_Q$.
It has been shown\cite{rf:29,rf:30} that the seesaw approximation for
the up-quark sector does not work effectively in the usual seesaw model.
However, owing to the factor 9 in Eq.~(\ref{eq:502}), which stems from
the USY, the seesaw approximation is justified in the present model.

The gauge eigenstates $f_L$ and $F_L$ and the mass eigenstates $f_L^{(M)}$
and $F_L^{(M)}$ are related by
\begin{equation}
   \left( \matrix{f_L^{(M)} \cr F_L^{(M)} \cr} \right) =
   \left( \matrix{U_f^\dagger & 0 \cr
                       0      & E \cr} \right)
   \left( \matrix{{\cal P}_f^\dagger & 0 \cr
                          0          & E \cr} \right)
   U_L^{f\dagger}
   \left( \matrix{f_L \cr F_L \cr} \right) ,
 \label{eq:513}
 \end{equation}
where $E$ is the $3\times 3$ unit matrix, and $(f, F) = (u, U)$ and $(d, D)$.
In the first seesaw approximation, the mass eigenstate $f_L^{(M)}$ of
left-handed ordinary quarks is given by
 \begin{equation}
    f_L^{(M)} \simeq U_f^\dagger {\cal P}_f^\dagger
                     (f_L - \rho_L^{f\dagger} F_L)
 \label{eq:514}
 \end{equation}
in terms of the gauge eigenstates. Therefore the CKM matrix $V$\cite{rf:22}
is constructed to be
 \begin{equation}
    V = U_u^\dagger {\cal P}_u^\dagger {\cal P}_d U_d .
 \label{eq:515}
 \end{equation}

In the case where $\Delta_u = \Delta_d = 0$,
the eigenvalues $\to_j^q$ and the eigenvectors $\tv_j^q$ of matrices $\tO_q$
lose dependence on $\delta_{ij}^q$. Therefore, in such a case, the CKM matrix
$V$ depends only on the difference $\phi = \phi_u - \phi_d$, and it is allowed
to impose the condition $\phi_d = 0$.

To realize the Wolfenstein parametrization\cite{rf:24} of the CKM matrix,
which is convenient to determine the shape of the unitarity triangle,
we carry out the phase transformation so
that $V_{11}$, $V_{21}$, $V_{12}$, $V_{32}$ and $V_{33}$ are real
($V_{11}, V_{12}, V_{33} > 0$ and $V_{21}, V_{32} < 0$). As a result,
we can obtain the Wolfenstein parameters $\rho$ and $\eta$ as
 \begin{equation}
    \rho \simeq -\displaystyle{\Re (V_{12}^\ast V_{32} V_{33}^\ast V_{13})
                         \over |V_{12}| |V_{21}| |V_{23}| |V_{32}|}, \qquad
    \eta \simeq \displaystyle{J \over |V_{12}| |V_{21}| |V_{23}| |V_{32}|},
 \label{eq:516}
 \end{equation}
where $J$ is the rephasing invariant Jarlskog parameter\cite{rf:23}:
 \begin{equation}
    J = \Im (V_{23} V_{12} V_{22}^\ast V_{13}^\ast) .
 \label{eq:517}
 \end{equation}

In the next section, we report on the numerical analysis of observable
quantities using the solutions in Eqs.~(\ref{eq:506}), (\ref{eq:507})
and (\ref{eq:510}) for the mass eigenvalue problems. To extract physically
meaningful analytical relations, these solutions are still too complicated.
For this purpose it is therefore reasonable to use approximate solutions,
as was done in the previous article\cite{rf:21}.  Here it is worthwhile to
emphasize that we can use all the results obtained there on the quark mass
differences, the sum rules among the absolute values of the CKM matrix, 
the analytic expressions for the Jarlskog and Wolfenstein parameters
in the situation considered presently.

\section{Numerical analysis}

The world averages of the absolute values of the CKM matrix elements are
estimated by the particle data group\cite{rf:32} as follows:
 \begin{equation}
   \left( \matrix{0.9745 \sim 0.9760 & 0.217 \sim 0.224 &
                  0.0018 \sim 0.0045  \cr
                  0.217  \sim 0.224  & 0.9737 \sim 0.9753 &
                  0.036  \sim 0.042   \cr
                  0.004  \sim 0.013 & 0.035 \sim 0.042 &
                  0.9991 \sim 0.9994  \cr} \right) .
 \end{equation}
Because the observed CKM matrix elements are given at $m_Z$, it is necessary to
know the values of running quark masses at $m_Z$. At the level of the 2-loop
renormalization group, Fusaoka and Koide\cite{rf:33} obtained the values of
running quark masses at $m_Z$ as
 \begin{eqnarray}
   & & m_u = 2.33^{+0.42}_{-0.45}\,{\rm MeV},  \quad
       m_c = 677^{+56}_{-61}\,{\rm MeV},  \quad
       m_t = 181 \pm 13 \,{\rm GeV},  \nonumber \\
       \noalign{\vskip0.2cm}
   & & m_d = 4.69^{+0.60}_{-0.66}\,{\rm MeV},  \quad
       m_s = 93.4^{+11.8}_{-13.0}\,{\rm MeV},  \quad
       m_b = 3.00 \pm 0.11 \,{\rm GeV}.
 \end{eqnarray}

Using the solutions in Eqs.~(\ref{eq:506}), (\ref{eq:507}) and
(\ref{eq:510}) for the mass eigenvalue problems, we proceeded with
the numerical analysis and found a set of parameters which reproduces
the experimental values of both the quark masses and the absolute values of
CKM matrix elements.  A typical solution obtained is as follows:
 \begin{equation}
             \left\{
 \begin{array}{lll}
   \phi_{11}^u = 0,  & \phi_{12}^u = 0.01455,  &
                       \phi_{13}^u = 0.07961,  \\
       \noalign{\vskip0.2cm}
   \phi_{21}^u = 0.002182,  & \phi_{22}^u = 0,  &
                             \phi_{23}^u = 0.09198,  \\
       \noalign{\vskip0.2cm}
   \phi_{31}^u = 0.07509,  & \phi_{32}^u = -0.2360,  &
                             \phi_{33}^u = 0
 \end{array} \right.
 \label{phiu}
 \end{equation}
and
 \begin{equation}
             \left\{
 \begin{array}{lll}
   \phi_{11}^d = 0,  & \phi_{12}^d = -0.1697,  &
                       \phi_{13}^d = 0.1970,  \\
       \noalign{\vskip0.2cm}
   \phi_{21}^d = -0.06880,  & \phi_{22}^d = 0,  &
                              \phi_{23}^d = 0.09669,  \\
       \noalign{\vskip0.2cm}
   \phi_{31}^d = 0.4753,  & \phi_{32}^d = -0.4347,  &
                            \phi_{33}^d = 0.
 \end{array} \right.
 \label{phid}
 \end{equation}
Substituting these values into Eqs.~(\ref{eq:403}) $\sim$ (\ref{eq:408}), 
the values of the quasi-democratic mass matrix elements are estimated to be
 \begin{equation}
             \left\{
 \begin{array}{ll}
   a_3^u = 0.9999,  & \delta_{12}^u =  0,  \\
       \noalign{\vskip0.2cm}
   a_2^u = 0.9912,  & \delta_{31}^u = -0.085,  \\
       \noalign{\vskip0.2cm}
   a_1^u = 0.9921,  & \delta_{23}^u =  0.085,
 \end{array} \right.
 \label{au}
 \end{equation}
and
 \begin{equation}
             \left\{
 \begin{array}{ll}
   a_3^d = 0.9927,  & \delta_{12}^d = 0,  \\
       \noalign{\vskip0.2cm}
   a_2^d = 0.9450,  & \delta_{31}^d = 0,  \\
       \noalign{\vskip0.2cm}
   a_1^d = 0.9193,  & \delta_{23}^d = 0.
 \end{array} \right.
 \label{ad}
 \end{equation}
With this set of parameters, we obtain the estimates
 \begin{equation}
   |V| =
   \left( \matrix{0.9753 & 0.2209 & 0.00357  \cr
                  0.2207 & 0.9745 & 0.04087 \cr
                  0.00869 & 0.04009 & 0.9992 \cr} \right)
 \end{equation}
for the absolute values of the CKM matrix elements,
 \begin{equation}
   \rho = 0.1213, \qquad \eta = 0.3765
 \end{equation}
for the Wolfenstein parameters, and
 \begin{equation}
   J = 3.008 \times 10^{-5}
 \end{equation}
for the Jarlskog parameter.

Provided that $M_u = 1.817 \times 10^2~{\rm GeV}$ and $M_d = 3.098~{\rm GeV}$,
the values of the running quark masses at $m_Z$ are obtained as
 \begin{eqnarray}
   & & m_u = 2.33~{\rm MeV}, \quad
       m_c = 6.77 \times 10^2~{\rm MeV}, \quad
       m_t = 1.81 \times 10^2~{\rm GeV},  \nonumber \\
       \noalign{\vskip0.2cm}
   & & m_d = 4.69~{\rm MeV}, \quad
       m_s = 9.34 \times 10~{\rm MeV}, \quad
       m_b = 3.00~{\rm GeV},
 \end{eqnarray}
using Eq.~(\ref{eq:510}).  All these results are consistent with
experimental results\cite{rf:32,rf:33,rf:34}.

\section{Discussion}

We have formulated the universal seesaw mechanism with the universal
strength for Yukawa couplings in the left-right symmetric
gauge group $G$, obtaining effectively the quasi-democratic mass matrices
of specific type. The left-right symmetric pairs of Higgs fields induce
interactions between the $i$-th generation of the light fermion doublets
and the $j$-th generation of the heavy fermion singlets with the coupling
constants $Ye^{i\phi_{ij}^q}$.
The universality of the strength for Yukawa couplings is taken strictly
here in the sense that $Y$ is common to the up- and down-quark sectors
in a left-right symmetric manner. 

The scale difference for the up- and down-quark sectors, $M_u$ and $M_d$, 
is ascribed to that of the heavy quark masses, $m_U$ and $m_D$, as
 \begin{equation}
   M_u \ : \  M_d \ \  \simeq \ \  {1 \over m_U} \ :\  {1 \over m_D}.
 \end{equation}
The departure of $M_u$ and $M_d$ from the electroweak scale $w_L$ is explained
by the seesaw factor $9Y^2 w_R / m_Q$. It is the relative phases $\phi_{ij}^q$
of the Yukawa couplings that explain generational variations of masses in
each quark sector. We found a set of values for the phase
parameters $\phi_{ij}^q$ which reproduces the experimental values for
the quark masses, the absolute values of CKM matrix elements, the Jarlskog
parameter and the Wolfenstein parameters.

Using Eq.~(\ref{eq:512}), we obtain
 \begin{equation}
    9Y^2 {w_R \over m_U} \simeq {m_t \over w_L} \sim 1 , \qquad
    9Y^2 {w_R \over m_D} \simeq {m_b \over w_L} \sim 0.02 ,
 \label{estimation}
 \end{equation}
from which the ratio of the bare masses of exotic quarks is estimated as
 \begin{equation}
    {m_D \over m_U} \sim 60 .
 \end{equation}
Equations~(\ref{seesawcondition}) and (\ref{estimation}) impose the condition
$81 Y^2 \gg 1$ on the strength $Y$ of the Yukawa coupling constants
for the first seesaw approximation to hold.  This implies that for
Yukawa coupling constants with strength $Y$ of the order, say, approximately
$\half$, the universal seesaw approximation is safely applicable to
all the sectors in our model.

In Eqs.~(\ref{phiu}) $\sim$ (\ref{ad}) we have given explicitly a set of
parameters which reproduces the experimental data. The values of the
parameters in Eqs.~(\ref{au}) and (\ref{ad}) display systematic departures
from the democratic limit $a_i^q = 1$ and $\delta_{ij}^q = 0$. It is not
possible to find, however, any order in Eqs.~(\ref{phiu}) and (\ref{phid})
for the values of the phase parameters $\phi_{ij}^q$. Therefore, as far as the
present set of parameters is concerned, it is difficult to conclude that there
is any indication of order or symmetry hidden in the relative phase couplings.
Further numerical analysis must be done to look for all the possible sets
of parameters which can reproduce the present experimental results.

In this article the arguments were restricted to the quark sector. The present
formalism will be applied to the lepton sector, and the puzzles of solar
and atomspheric neutrinos will be analized in a future publication.


\begin{thebibliography}{99}
\bibitem{rf:1}
  H.~Fritzsch, Phys. Lett. {\bf 73B} (1978), 317; {\bf 85B} (1979), 81;
               Nucl. Phys. {\bf B155} (1979), 189. \\
  H.~Fritzsch and Zhi-zhong~Xing, Phys. Lett. {\bf B353} (1995), 114.
\bibitem{rf:2}
  B.~Stech, Phys. Lett. {\bf 130B} (1983), 189.
\bibitem{rf:3}
  M.~Shin, Phys. Lett. {\bf 145B} (1984), 285.
\bibitem{rf:4}
  M.~Gronau, R.~Johnson and J.~Schechter, 
             Phys. Rev. Lett. {\bf 54} (1985), 2176.
\bibitem{rf:5}
  S.~N.~Gupta and J.~M.~Johnson, Phys. Rev. {\bf D44} (1991), 2110.
\bibitem{rf:6}
  G.~C.~Branco, L.~Lavoura and F.~Mota, Phys. Rev. {\bf D39} (1989), 3443. \\
  G.~C.~Branco and J.~I.~Silva-Marcos, Phys. Lett. {\bf B331} (1994), 390. \\
  Y.~Koide, Mod. Phys. Lett. {\bf A12} (1997), 2655.
\bibitem{rf:7}
  H.~Harari, H.~Haut and J.~Weyers, Phys. Lett. {\bf 78B} (1978), 459.
\bibitem{rf:8}
  T.~Goldman and G.~J.~Stephenson,~Jr., Phys. Rev. {\bf D24} (1981), 236.
\bibitem{rf:9}
  Y.~Koide, Phys. Rev. Lett. {\bf 47} (1981), 1241; 
            Phys. Rev. {\bf D28} (1983), 252; {\bf D39} (1989), 1391.
\bibitem{rf:10}
  P.~Kaus and S.~Meshkov, Mod. Phys. Lett. {\bf A3} (1988), 1251;
                          Phys. Rev. {\bf D42} (1990), 1863.
\bibitem{rf:11}
  L.~Lavoura, Phys. Lett. {\bf B228} (1989), 245.
\bibitem{rf:12}
  M.~Tanimoto, Phys. Rev. {\bf D41} (1990), 1586.
\bibitem{rf:13}
  H.~Fritzsch and J.~Plankl, Phys. Lett. {\bf B237} (1990), 451.
\bibitem{rf:14}
  Y.~Nambu, in {\it Proceedings of the International Workshop on Electroweak
  Symmetry Breaking}, Hiroshima (World Scientific, Singapore, 1992), p.1.
\bibitem{rf:15}
  G.~C.~Branco, J.~I.~Silva-Marcos and M.~N.~Rebelo, 
                                   Phys. Lett. {\bf B237} (1990), 446.
\bibitem{rf:16}
  G.~C.~Branco and J.~I.~Silva-Marcos, Phys. Lett. {\bf B359} (1995), 166. \\
  G.~C.~Branco, D.~Emmanuel-Costa and J.~I.~Silva-Marcos, 
                Phys. Rev. {\bf D56} (1997), 107.
\bibitem{rf:17}
  J.~Kalinowski and M.~Olechowski, Phys. Lett. {\bf B251} (1990), 584.
\bibitem{rf:18}
  P.~M.~Fishbane and P.~Kaus, Phys. Rev. {\bf D49} (1994), 3612; 4780; 
                              Z. Phys. {\bf C75} (1997), 1.
\bibitem{rf:19}
  P.~M.~Fishbane and P.~Q.~Hung, Mod. Phys. Lett. {\bf A12} (1997), 1737; 
                                 Phys. Rev. {\bf D57} (1998), 2743.
\bibitem{rf:20}
  T.~Teshima and T.~Sakai, Prog. Theor. Phys. {\bf 97} (1997), 653.
\bibitem{rf:21}
  I.~S.~Sogami, K.~Nishida, H.~Tanaka
            and T.~Shinohara, Prog. Theor. Phys. {\bf 99} (1998), 281.
\bibitem{rf:22}
  N.~Cabibbo, Phys. Rev. Lett. {\bf 10} (1963), 531. \\
  M.~Kobayashi and T.~Maskawa, Prog. Theor. Phys. {\bf 49} (1973), 652.
\bibitem{rf:23}
  C.~Jarlskog, Phys. Rev. Lett. {\bf 55} (1985), 1039; 
               Z. Phys. {\bf C29} (1985), 491. \\
  I.~Dunietz, O.~W.~Greenberg and Dan-di~Wu, 
              Phys. Rev. Lett. {\bf 55} (1985), 2935.
\bibitem{rf:24}
  L.~Wolfenstein, Phys. Rev. Lett. {\bf 51} (1983),1945.
\bibitem{rf:25}
  Z.~G.~Berezhiani, Phys. Lett. {\bf 129B} (1983), 99; {\bf 150B} (1985), 177; 
                    Yadern. Fiz. {\bf 42} (1985), 1309. \\
  D.~Chang and R.~N.~Mohapatra, Phys. Rev. Lett. {\bf 58} (1987), 1600. \\
  S.~Rajpoot, Phys. Lett. {\bf B191} (1987), 122; 
              Phys. Rev. {\bf D36} (1987), 1479. \\
  A.~Davidson and K.~C.~Wali, Phys. Rev. Lett. {\bf 59} (1987), 393; 
                                               {\bf 60} (1988), 1813. \\
  A.~Davidson, S.~Ranfone and K.~C.~Wali, Phys. Rev. {\bf D41} (1990), 208. \\
  S.~Ranfone, Phys. Rev. {\bf D42} (1990), 3819.
\bibitem{rf:26}
  I.~S.~Sogami, Prog. Theor. Phys. {\bf 85} (1991), 141. \\
  I.~S.~Sogami and T.~Shinohara, Prog. Theor. Phys. {\bf 86} (1991), 1031. \\
  T.~Shinohara, Prog. Theor. Phys. {\bf 90} (1993), 1057.
\bibitem{rf:27}
  I.~S.~Sogami, T.~Shinohara and Y.~Egawa, 
                 Prog. Theor. Phys. {\bf 87} (1992), 1005.
\bibitem{rf:28}
  I.~S.~Sogami and T.~Shinohara, Phys. Rev. {\bf D47} (1993), 2905.
\bibitem{rf:29}
  Y.~Koide and H.~Fusaoka, Z. Phys. {\bf C71} (1996), 459.
\bibitem{rf:30}
  Y.~Koide and M.~Tanimoto, Z. Phys. {\bf C72} (1996), 333. \\
  Y.~Koide and H.~Fusaoka, Prog. Theor. Phys. {\bf 97} (1997), 459. \\
  Y.~Koide, Phys. Rev. {\bf D56} (1997), 2656. \\
  T.~Morozumi, T.~Satou, M.~N.~Rebelo and M.~Tanimoto, 
                         Phys. Lett. {\bf B410} (1997), 233.
\bibitem{rf:31}
  G.~Senjanovic and R.~N.~Mohapatra, Phys. Rev. {\bf D12} (1975), 1502. \\
  G.~Senjanovi\'{c}, Nucl. Phys. {\bf B153} (1979), 334. \\
  R.~N.~Mohapatra and G.~Senjanovi\'{c},
   Phys. Rev. Lett. {\bf 44} (1980), 912; Phys. Rev. {\bf D23} (1981), 165.
\bibitem{rf:32}
  Particle Data Group, C.~Caso~et~al., Eur. Phys. J. {\bf C3} (1998), 1.
\bibitem{rf:33}
  H.~Fusaoka and Y.~Koide, Phys. Rev. {\bf D57} (1998), 3986.
\bibitem{rf:34}
  M.~Neubert, Int. J. Mod. Phys. {\bf A11} (1996), 4173.
\end{thebibliography}
\end{document}